\documentclass[aps,prb,reprint,groupedaddress,superscriptaddress,floatfix]{revtex4-2}

\usepackage{graphicx}

\usepackage[inline]{enumitem}
\setlist[enumerate]{label=\roman*.}

\usepackage[version=4]{mhchem}
\usepackage[separate-uncertainty = true,multi-part-units=single]{siunitx}
\DeclareSIUnit\angstrom{\text{Å}}

\usepackage{cleveref}
\crefname{figure}{Fig.}{Figures}
\crefname{table}{Table}{Tables}

\usepackage{xcolor}

\let\oldsim\sim
\renewcommand{\sim}{\mathord{\oldsim}}

\newcommand{\SiTe}{\ce{Si2Te3}}

\begin{document}

\title{High Pressure Structural Behavior of Silicon Telluride (\SiTe{}) Nanoplates}

\author{Bohan Li}
\affiliation{%
Department of Physics and Astronomy, University of California Davis,  Davis, CA 95616, USA
}
\affiliation{%
Department of Chemistry, University of California Davis,  Davis, CA 95616, USA
}

\author{Frank Cerasoli}
\affiliation{%
Department of Chemistry, University of California Davis,  Davis, CA 95616, USA
}%

\author{Ethan Chen}
\affiliation{%
Department of Chemistry, University of California Davis,  Davis, CA 95616, USA
}%

\author{Martin Kunz}
\affiliation{%
Advanced Light Source, Lawrence Berkeley National Laboratory, Berkeley, CA, USA
}%

\author{Davide Donadio}
\affiliation{%
Department of Chemistry, University of California Davis,  Davis, CA 95616, USA
}%

\author{Kristie J. Koski}
\email{koski@ucdavis.edu}
\affiliation{%
Department of Chemistry, University of California Davis,  Davis, CA 95616, USA
}%

\date{\today}

\begin{abstract}
The high-pressure behavior of silicon telluride (\SiTe{}), a two-dimensional (2D) layered material, was investigated using synchrotron X-ray powder diffraction in a diamond anvil cell to \SI{11.5}{GPa} coupled with first-principles theory. \SiTe{} undergoes a phase transition at $<\SI{1}{GPa}$ from a trigonal to a hexagonal crystal structure. At higher pressures ($> \SI{8.5}{GPa}$), X-ray diffraction showed the appearance of new peaks possibly coincident with a new phase transition, though we suspect \SiTe{} retains a hexagonal structure. Density functional theory calculations of the band structure reveal metallization above \SI{9.1}{GPa} consistent with previous measurements of the Raman spectra and disappearance of color and transparency at pressure. The theoretical Raman spectra reproduce the prominent features of the experiment, though a deeper analysis suggests that the orientation of Si dimers dramatically influences the vibrational response. Given the complex structure of \SiTe{}, simulation of the resulting high-pressure phase is complicated by disordered vacancies and the initial orientations of Si--Si dimers in the crushed layered phase.

\end{abstract}

\maketitle

\section{Introduction}

Silicon telluride (\SiTe{}) is a brilliant red, transparent, two-dimensional (2D) layered material that offers promise for its optoelectronic properties \cite{keuleyan2015silicon, wangChemicallyTunableFull2018, wu2017structure, chen2019probing}, intercalation capabilities \cite{keuleyan2015silicon, wu2018morphology,wu2018resistive,huynh2023hafnium}, and because it is uniquely compatible with current silicon semiconductor processing technologies. \SiTe{} exhibits photoluminescence from defect trap states around \SI{650}{nm} offering promise as a photomaterial \cite{wangChemicallyTunableFull2018, ziegler1977photoelectric}.
Measurements show possible evidence of memristor behavior \cite{wu2018resistive}, and recent calculations suggest that \SiTe{} is a superior thermoelectric with an unprecedented ZT of 1.86 \cite{juneja2017high}.

From a fundamental perspective, \SiTe{} shows a unique thermodynamic phase behavior. \textcite{johnson2019pressure} demonstrated that \SiTe{} likely undergoes a possible semiconductor-to-metal electronic phase transition under pressure at \SI{9.5 \pm 0.5}{GPa} using Raman scattering of single plates in a diamond anvil cell.
Optical measurements showed that silicon telluride went from transparent red to opaque grey/black under pressure.
Further, Raman modes vanished at high pressures suggesting a change in the polarizability of the crystal.
However, these studies lacked structural investigations to identify any accompanying phase transitions.
Despite a wealth of novel properties, the fundamental thermodynamic phase behavior of silicon telluride remains largely unknown.
Calculations by \textcite{bhattaraiPressureInducedInsulatorMetal2021} suggested that compression beyond \SI{7}{GPa} induces a pressure-driven phase transition from a semiconductor to a metal with a coincident structural change.
Other calculations also suggest that silicon telluride compressed in a silicon-based pressure medium could lead to the formation of SiTe \cite{steinberg2016search}, a long-sought phase that is expected to exhibit topological insulator properties and is impossible to achieve under ambient pressures \cite{ma2016proposed, wang2018high}.
The change in stoichiometry, though, seems unsupported by experiment. 
\textcite{grzechnik2022chemical} investigated pressure-dependent phases of silicon telluride by mixing powders of Si and Te and heating them at high pressures and temperatures in a large volume multi-anvil press while performing \emph{in situ} synchrotron X-ray diffraction (XRD).
Several unique phases that can be formed from mixtures of silicon and tellurium were identified \cite{grzechnik2022chemical}.
They also measured bulk synthesized \SiTe{} as a function of pressure and temperature.
At elevated temperatures, \textcite{grzechnik2022chemical} found several different phases from silicon and tellurium including a \ce{Mn5Si3} hexagonal type phase and several clathrates.
In the same article, bulk-grown \SiTe{} was compressed at ambient temperature and it was found that above \SI{12}{GPa}, \SiTe{} undergoes a phase change to amorphous; however, no data of these measurements was provided to adequately evaluate these conclusions.

In this joint experimental and computational investigation, we use a diamond anvil cell to hydrostatically compress vapor-phase grown \SiTe{} nanoplates to high pressures at ambient temperature.
\emph{In situ} synchroton X-ray diffraction at pressure reveals that \SiTe{} undergoes a phase transition at fairly low pressures ($\sim\SI{0.2}{GPa}$), corresponding to a structural change from trigonal ($P\overline{3}1c$) to hexagonal with a higher symmetry ($P6/mmm$).
Peaks associated with the van der Waals (vdW) gap vanish as expected in layered materials \cite{grochala2007chemical}.
At higher pressures ($>\SI{8.5}{GPa}$), X-ray diffraction shows new peaks that cannot be indexed with any certainty as a hexagonal crystal.
No amorphization is seen to \SI{11.5}{GPa}, and returning to ambient pressures does not restore the original structure.
\emph{In-situ} XRD and Raman spectroscopy measurements are corroborated with density functional theory (DFT) calculations, identifying a semiconductor-to-metal phase transition in \SiTe{} \cite{johnson2019pressure}.
Our calculations reveal a structural phase transition that eliminates the vdW gap, as would be expected with increased pressures, accompanied by an increase in covalent bond count between silicon atoms and a change in bond coordination from tetrahedral to octahedral.
Furthermore, this transition is irreversible and the interlayer covalent bonds remain intact when pressure is released.
Our calculations predict that a narrow electronic band gap and a characteristic Raman peak are recovered upon releasing the high pressure, which supports experimental evidence of both the electronic and structural transitions. 
Disorder in the bulk phase presents challenges when trying to resolve the atomic structure with XRD, which is clarified throughout the text.

Our experiments and calculations indicate that the stoichiometry is conserved under single-crystal compression.
Further, the electronic phase transition is reversible returning from a metal to a semiconductor but structural phase transitions are irreversible and associated with release to a more disordered phase not easily indexed with XRD.

\begin{figure*}
    \centering
    \includegraphics{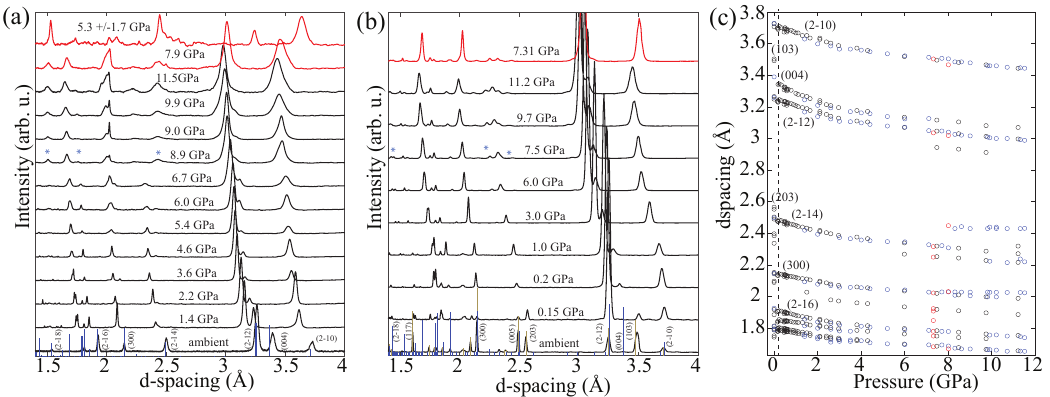}
    \caption{(a) Stacked plot of \SiTe{} as a function of pressure. Red plots are decreasing pressure. (b) Additional experimental pressure run of \SiTe{} where sample orientation shows other indices. Asterisks (*) denote new peaks at higher pressures. (c) d-spacing as a function of pressure. Black, $\circ$,  and blue circles, $\color{blue} \circ$, are two different runs. Red circles, $\color{red} \circ$, are decreasing pressure. }
    \label{Fig1}
\end{figure*}

\section{Methods}

\subsection{Experimental}

\subsubsection{Synthesis}
Silicon telluride was synthesized using previously developed vapor-phase methods \cite{keuleyan2015silicon}.
Briefly, a ceramic boat containing a mixture of \SI{0.18}{g} Te and \SI{0.07}{g} Si was placed in the center of a \SI{25.4}{mm} quartz tube.
Upstream, a ceramic boat containing \SI{0.10}{g} Te was placed.
Downstream, at \qtyrange[range-phrase=--]{12}{15}{cm} from the center, a quartz substrate was used for the large area vapor-liquid-solid (VLS) growth.
The furnace was heated to \SI{800}{\celsius} at a ramp rate of \SI{60}{\celsius} per minute.
Once the furnace reached \SI{800}{\celsius}, the furnace was slowly cooled to room temperature.
Plates were stored in separate containers under inert \ce{N2} in a glovebox.

\subsubsection{Characterization}
Powder X-ray diffraction of \SiTe{} at ambient conditions was collected on a Bruker Eco Advance with a Cu $\text{K}_{\alpha}$ source ($\lambda = \SI{1.54}{\angstrom}$).
TEM images and electron diffraction (SAED) were collected on a JEOL 2100fac. Raman spectra were acquired on a home-built system with an inverted Leica DMi8, Princeton Instruments Isoplane SC320 with an 1800 groove/mm grating, Pixis CCD Camera, and Coherent Sapphire SF laser at $\lambda = \SI{532.15}{nm}$ at $\sim \SI{5}{mW}$ laser power.
A Semrock RazorEdge filter and dichroic give a cutoff of $\sim 70\text{--}\SI{100}{cm^{-1}}$.
Acquisition times were 5 seconds with 5 spectral averages.
Ultraviolet (UPS) was collected on the Kratos Axis Supra using a He-I (\SI{21.22}{eV}) source with an applied bias of \SI{-9}{V}.
Optical images were acquired on a Leica M205C.
Characterization of \ce{Si2Te3} including optical images, X-ray photoelectron spectra, and scanning electron microscopy energy dispersive characterization of the elemental constituency can be found in Fig. S1 in the supporting information (SI). 

\subsubsection{High-Pressure X-ray Diffraction Measurements}
A gas membrane-driven symmetric diamond anvil cell from Deutsches Elektronen-Synchrotron (DESY)  with \SI{400}{\um} diamond culets was used for high-pressure measurements.
Spring-steel gaskets were pre-indented to $\sim\SI{100}{\um}$ thickness and a centered \SI{200}{\um} diameter gasket hole was drilled with a Cameron Micro Drill Press Series 214-D3.
Several grains of annealed ruby spheres from Alamax were placed into the gasket hole to measure the pressure inside the diamond cell. Ethylcyclohexane (ECH) was used as a pressure-transmitting medium.
Loading the sample was done using an acupuncture needle to scrape the \SiTe{} growth substrate and place the platelets into the DAC.
A drop of ECH was placed onto the substrate; the substrate was scratched with the needle and then drawn back up into the syringe and dropped into the DAC which was quickly closed.
The sample loading is done in less than a minute from exposure of the \SiTe{} to complete loading to prevent oxidation of the \SiTe{}.
ECH is not a very hydrostatic pressure medium but other media either oxidize the sample, like $4:1$ methanol ethanol or silicone oil or are not as easy to load with \SiTe{}, like $1:1$ $\text{pentane}:\text{isopentane}$.
Ethylcyclohexane has been used in several diamond anvil cell investigations previously and maintains somewhat hydrostatic conditions to about \SI{4}{GPa} \cite{koski2008structural}.

High-pressure X-ray diffraction patterns (XRD) were acquired on high-pressure beamline 12.2.2 of the Advanced Light Source using an energy of $\sim \SI{25}{keV}$ ($\lambda = \SI{0.496}{\angstrom}$) \cite{kunz2005beamline}.
XRD was acquired using a Pilatus 2 M detector. Typical acquisition times were between 2--10 minutes.
The sample to detector distance and detector tilt were calibrated with a \ce{CeO2} NIST standard.
All 2D XRD images were integrated into 1D patterns using Dioptas \cite{prescher2015dioptas}.
The pressure was determined using the ruby fluorescence method.
About 2--3 ruby spheres were placed into the cell and used to determine the pressure.
Ruby fluorescence was acquired using an online ruby system with a \SI{488}{nm} laser to determine the pressure \cite{Ruby1,Ruby2}. 

XRD pattern indexing was performed using the DICVOL program \cite{boultif2004powder} and Treor90 \cite{werner1964trial,werner1985treor} implemented in FULLPROF \cite{rodriguez2001fullprof} through MATCH \cite{putz2015match}.
Structure solutions were also determined using GSAS-II \cite{toby2013gsas}.
Lattice parameters were confirmed via LeBail analysis using FOX (Free Objects for Crystallography) \cite{favre2002fox}.
Above \SI{0.2}{GPa}, DICVOL assigns a smaller unit cell than expected, giving a unit cell that is $a / \sqrt{3} \times c/2$ while FOX gives the expected $a \times c$.   

\subsection{Theoretical}

\subsubsection{Computational Details}
\label{computational-details}

First-principles electronic structure DFT calculations are performed using \textsc{Quantum ESPRESSO} (QE) v7.2 \cite{QE-2009,QE-2017,doi:10.1063/5.0005082}.
To appropriately model layered \SiTe{}, we choose an exchange-correlation functional that contains an additional nonlocal term to describe van der Waals interactions (vdW-DF-cx)\cite{thonhauserVanWaalsDensity2007,thonhauserSpinSignatureNonlocal2015,langrethDensityFunctionalSparse2009,berlandVanWaalsForces2015,sabatiniStructuralEvolutionAmino2012}.
The \SiTe{} crystal geometry is obtained from \textcite{ploog1976crystal} and was used to conduct our DFT simulations. 
Atomic sites are described by projector augmented wave (PAW) pseudopotentials from the PSLibrary \cite{dalcorsoPseudopotentialsPeriodicTable2014}. 
Plane waves are expanded up to a kinetic energy cutoff of \SI{60}{Ry} (\SI{240}{Ry}) for wavefunctions (charge density), and the self-consistent field (SCF) convergence threshold is set to \SI{E-8}{Ry}.
A $\Gamma$-centered Monkhorst-Pack mesh is sampled with a resolution of $6 \times 6 \times 6$.
Electronic densities of states (DOS) are computed from a non-SCF diagonalization on a $12 \times 12 \times 12$ k-points grid centered at the $\Gamma$-point with occupations calculated with the tetrahedron method \cite{blochlImprovedTetrahedronMethod1994}.

\subsubsection{Raman and Phonon Spectra}

Raman spectra are computed by density functional perturbation theory (DFPT) and the N+1 theorem implemented in the PHonon code within QE\cite{baroni_phonons_2001,lazzeri_first-principles_2003}.
To compute the Raman spectra we used the local-density approximation (LDA) exchange-correlation functional and Norm-Conserving (NC) pseudo potentials, as other computational frameworks are not implemented.
LDA pseudopotentials were also obtained from the PSLibrary \cite{dalcorsoPseudopotentialsPeriodicTable2014}.
In Raman spectra calculations under this parameterization, kinetic energy cutoffs are increased to \SI{80}{Ry} and \SI{320}{Ry} for the wavefunctions and charge density, which achieves an estimated accuracy of \SI{E-5}{Ry} for total energy.

LDA gives similar structural and vibrational properties as the vdW-DF-cx functional for other layered chalcogenides, e.g., MoS$_2$\cite{chen_strongly_2019}.
The structures previously obtained with PAW pseudo potentials are optimized to equilibrium with NC pseudo potentials by relaxing atomic positions until all forces converge below \num{E-5}.
Structural reorganization is minor and the resulting atomic displacements do not exceed \SI{0.1}{\angstrom}. 
Non-resonant Raman coefficients and macroscopic dielectric constants are computed at the $\Gamma$ point.
SCF energy is converged to \SI{E-8}{Ry}, and the acoustic sum rule is imposed on the dynamical matrix together with nonanalytical corrections to facilitate LO--TO splitting near $\Gamma$.
Using the calculated interatomic force constants and phonon frequencies, phonon polarizations are obtained as the eigenvectors of the dynamical matrix and visualized with the Interactive Phonon Visualizer \cite{phononVisualizer}. 

\section{Results and Discussion}

\subsection{Experimental}

X-ray diffraction of \SiTe{} was collected under high pressures in the diamond anvil cell at beamline 12.2.2 of the Advanced Light Source and is shown in Fig.~\ref{Fig1}.
Peaks at ambient pressure were indexed using MATCH \cite{putz2015match} and assigned according to the crystal structure determined by \textcite{ploog1976crystal}.
At initial ambient pressure, but loaded into the diamond cell under ethylcyclohexane, \SiTe{} has a trigonal crystal structure (space group: $P\overline{3}1c$) with lattice constants $a = \SI{7.44}{\angstrom}$ and $c = \SI{12.45}{\angstrom}$ with a volume of \SI{597.4}{\angstrom^3}.
This is in contrast to \SiTe{} outside the diamond anvil cell where we found that $a = \SI{7.42}{\angstrom}$, $c = \SI{13.49}{\angstrom}$ and a volume of \SI{641.5}{\angstrom^3} suggesting that simply putting the \SiTe{} in the cell under ethylcyclohexane substantially affects the initial structure.
With pressure, the (004) peak compresses into the ($2 \bar{1} 2$) peak suggesting compression of the vdW gap.
Rietveld refinement of the crystal to the trigonal phase identified by \textcite{ploog1976crystal} was used to determine the lattice constants below \SI{0.23}{GPa}.
The trigonal structure remains until only \SI{0.23}{GPa}, where the structure solution identifies the crystal structure as hexagonal (space group: $P6/mmm$).
Thus, with moderate pressure, \SiTe{} nanoplates are in a higher symmetry phase.
At \SI{1}{GPa}, we find that that \SiTe{} has lattice constants of $a = \SI{7.40}{\angstrom}$ $c = \SI{13.17}{\angstrom}$ and a volume of \SI{605.9}{\angstrom^3}.
New peaks appear around 1.5, 2.1 and \SI{2.6}{\angstrom} between \SI{6.5}{GPa} and \SI{9}{GPa} while other peaks near \SI{2.4}{\angstrom} vanish including a multiplet around \SI{2.3}{\angstrom}.
The XRD pattern can be indexed with monoclinic or hexagonal, but the structure solution does not achieve an acceptable figure of merit using DICVOL, GSAS, and FOX.
There are only about 12 defined peaks to extract a structure solution from and all peaks are broad.
Assuming a phase that is probably monoclinic, this is insufficient data for a proper crystal structure solution.
At \SI{11.2}{GPa}, with a hexagonal structure, we found lattice constants of $a = \SI{6.86}{\angstrom}$ and $c = \SI{12.23}{\angstrom}$ with a volume of \SI{498.4}{\angstrom^3}. 

\textcite{grzechnik2022chemical} compressed \SiTe{} at room temperature and reported that it becomes amorphous above \SI{12}{GPa} but provide no powder patterns of this transition or behavior of \SiTe{} leading up to this transition. 
Although we encounter peak broadening at higher pressures, we do not find any evidence of amorphization up to pressures of \SI{9.5 \pm 0.5}{GPa} above which the Raman modes disappear\cite{johnson2019pressure}.
There is no evidence for a change in stoichiometry either.  

Diffraction patterns were collected upon decompression of the diamond cell.
The cell, however, maintains pressure when released.
Opening the cell exposes the silicon telluride to air and hydrolyzes the samples.
XRD patterns associated with a decrease in pressures are shown in red (Fig.~\ref{Fig1}).
Upon decompression, the high-pressure XRD pattern is maintained until $\sim \SI{7}{GPa}$.
Below $\sim \SI{7}{GPa}$, \SiTe{} does not return to its original structure.

\begin{figure}
    \centering
    \includegraphics{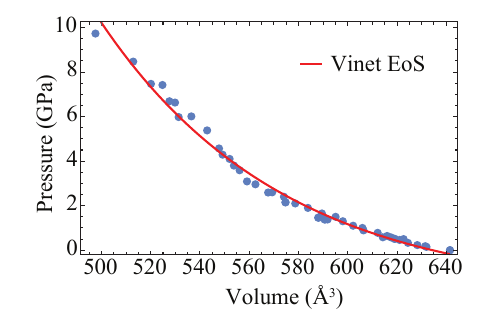}
    \caption{The Vinet equation of state fit to the volume change of \SiTe{} with pressure.}
    \label{Fig2}
\end{figure}

A Vinet equation of state \cite{vinet1987compressibility, jeanloz1988universal} was used to fit the XRD data.
A fit of the pressure vs experimental volume to a Vinet equation of state (Fig.~\ref{Fig2}) with the zero-pressure volume yields a bulk modulus of $K_o = \SI{16.7 \pm 1.7}{GPa}$, a pressure derivative of the bulk modulus of $K_o' = 8.2 \pm 0.7 $, and $V_o = \SI{634.4 \pm 2.9}{\angstrom^3}$ respectively. 

\subsection{Theoretical}

\subsubsection{Layered Phase Structure}

\begin{figure}[ht]
    \centering
    \includegraphics[width=\linewidth]{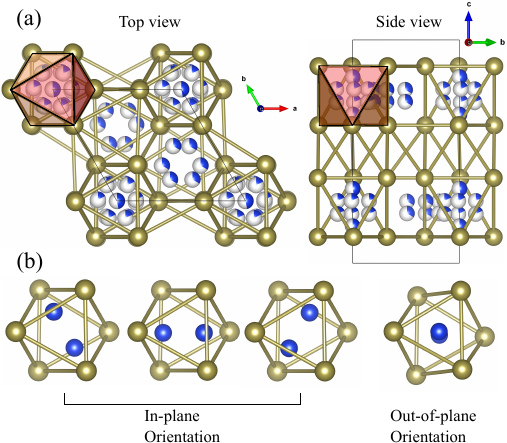}
    \caption{\SiTe{} structure. Atoms are colored yellow (blue) for Te (Si). The Crystallographic Information File for \ce{Si2Te3} is found on the Crystallography Open Database given by \textcite{ploog1976crystal}. (a) Top and side views of bulk \ce{Si2Te3} structures shows hexagonal close packed lattice. One of the Te octahedron cages is drawn out with edges and faces to make the octahedral geometry recognizable. The partial blue coloring of the Si atoms represents the partial occupation. (b) Four possible orientations including three in-plane orientations and an out-of-plane orientation for Si dumbbells inside a Te layer under the top view as (a).  For the out-of-plane orientation, the viewpoint is slightly tilted to show the Si atom beneath the top layer.}
    \label{fig:structure}
\end{figure}

The initial structure of \ce{Si2Te3} was determined from prior XRD experiments \cite{ploog1976crystal,keuleyan2015silicon,RevisitingDronskowski2017}.
At ambient pressures, \ce{Si2Te3} has a trigonal crystal structure with 12 Te atoms and 8 Si atoms in the conventional cell.
The Te atoms are arranged in a hexagonal geometry, as shown in \cref{fig:structure}(a).
\ce{Si2Te3} possesses a relatively complicated structure because pairs of Si atoms form dimers (dumbbells) that partially occupy various positions and orientations between the Te layers. 
Within each Te layer reside two unique site clusters that host one silicon dimer with probabilistic orientation, which results in a disordered crystal.
Two of the dumbbells have four possible orientations as shown in \cref{fig:structure}(b), and their \textit{measured} occupation factors are as follows: The first is an out-of-plane (vertical) orientation with 47\% occupation; the other three possible orientations are in-plane (horizontal), each occupied with equal probability of 18\%.
There is no out-of-plane orientation for the Si dumbbell for the Si clusters that are not vertically aligned between layers.
Instead, there are only three in-plane orientations, each with an equal probability of 33\% \cite{ploog1976crystal,shenVariabilityStructuralElectronic2016}. 

Partial occupancy among the sites forming silicon dimers presents a unique challenge for our theoretical analysis of \SiTe{}.
In the unit cell, there are 1001 possible configurations.
To tackle the complexity of partial occupation, we used a site permutation search (SPS) algorithm in a unit cell \cite{cerasoliEffectiveOptimizationAtomic2024a} that allows us to identify the \SiTe{} structure with the minimal energy.
Due to the relatively small number of possible configurations in a unit cell, the SPS algorithm is used to predict the formation energy of every configuration, utilizing the MEGNet pre-trained Crystal Graph Convolutional Neural Network (CGCNN), and identify the structure with the lowest energy \cite{chen_graph_2019}. 
Finally, the geometry was optimized by relaxing both cell parameters and atomic positions at \SI{0}{GPa} in QE.
The resulting structure is hereafter referred to as the ``layered phase".
Slight distortions from a hexagonal structure were incurred by freely optimizing the cell parameters.
The angle between the basal lattice vectors \textbf{a} and \textbf{b} increases to $122.1^\circ$, and the angle between \textbf{b} and \textbf{c} decreases to $89.4^\circ$.
We attribute this minor divergence from the hexagonal structure to our choice of a single orientation for each of the silicon dimers, which does not accurately capture the partial occupation observed experimentally.

\subsubsection{Layered Phase of \SiTe{} at Ambient Conditions}

\begin{figure}[b]
    \centering
    \includegraphics[width=\linewidth]{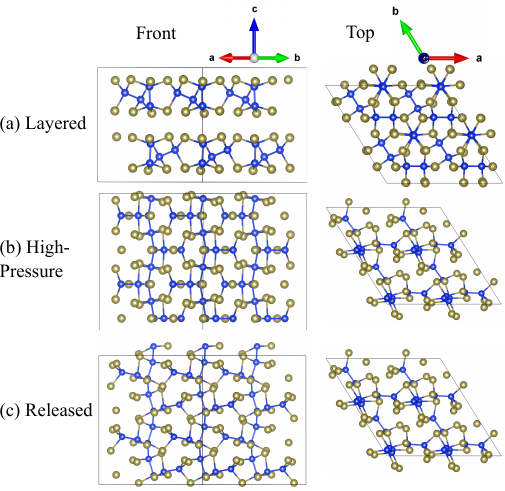}
    \caption{Structures of various phases. Layered phase (a) is shown as a $2 \times 2 \times 1$ supercell. High-pressure (b) and released (c) phases are shown as $2 \times 2 \times 2$ supercells.}
    \label{fig:phase-structure}
\end{figure}

The \ce{Si2Te3} ground state layered phase at \SI{0}{GPa} is shown in \cref{fig:phase-structure}(a). 
This configuration has \SI{50}{\percent} of the Si dumbbells aligned vertically and the other \SI{50}{\percent} aligned in the plane.
In this case, both layers in the unit cell have exactly one pair of vertical Si dumbbells.
The layered phase unit contains 24 Si--Te bonds and 4 Si--Si bonds with respective lengths of \SI{2.55}{\angstrom} and \SI{2.31}{\angstrom}.
Each Si site is tetrahedrally coordinated, forming bonds with three Te atoms and one other Si atom, while each Te atom bonds with two Si atoms.
The mean value of Si--Si--Te (Te--Si--Te) bond angles is \SI{106 \pm 3.7}{\degree} (\SI{112 \pm 5.88}{\degree}). 

\begin{figure*}
    \centering
    \includegraphics[width=\linewidth]{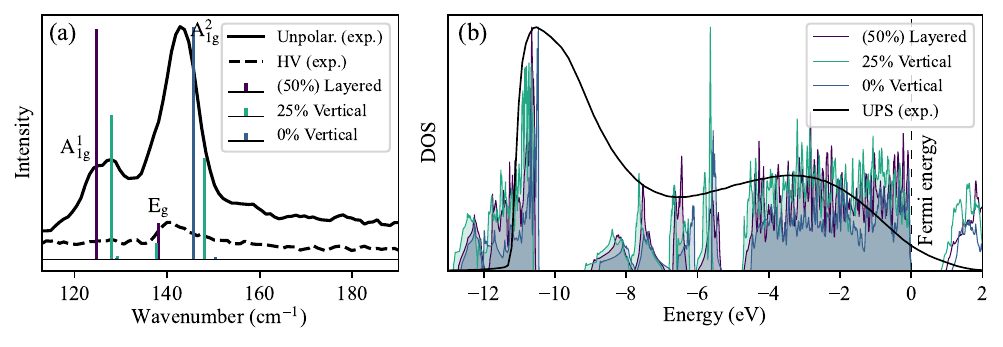}
    \caption{Properties of \SiTe{} system at the ambient state. (a) Theoretical Raman spectra with different Si dumbbell orientations compared with the experimental Raman spectra from \textcite{wangChemicallyTunableFull2018}. $\mathrm{A}_{\mathrm{1g}}^1$ and $\mathrm{A}_{\mathrm{1g}}^2$ are two out-of-plane phonon modes; $\mathrm{E}_{\mathrm{g}}$ is an in-plane phonon mode. (b) Density of states along with the experimental UPS. The zero point of energy is set at the Fermi level. The zero point is set by aligning the maximum of the UPS spectrum with that of the theoretical density of states for the layered phase. Unpolar.\@ (exp.) is the unpolarized Raman spectrum from the experiment. HV (exp.) is a perpendicular configuration with polarization that eliminates out-of-plane $\mathrm{A}_{\mathrm{1g}}$ modes. UPS (exp.) is the UPS spectrum from the experiment. 
    }
    \label{fig:layered-phase-properties}
\end{figure*}

The layered phase is a semiconductor with an indirect band gap of \SI{1.01}{eV}, as shown in Fig.~S4(a) and \cref{fig:raman-compare-dos}(c), which is very close to the experimental indirect band gap of \SI{0.98}{eV}, measured at ambient conditions \cite{wangChemicallyTunableFull2018}.
The direct band gap is not accurately captured by DFT.
Other studies have employed the GW and Bethe-Salpeter Equation (BSE) methods to refine predictions of the direct band gap in layered \SiTe{} \cite{bhattaraiAnisotropicOpticalProperties2020,chenAnisotropicOpticalProperties2020}.
The calculated density of states (DOS) for the layered phase agrees well with the experimental ultraviolet photoelectron spectroscopy (UPS) spectrum as shown in \cref{fig:layered-phase-properties}(b), where the Fermi energy is aligned to \SI{0}{eV} and the highest peaks of the DOS and UPS spectrum are scaled to the same magnitude. 

Although our DFT calculations of the layered phase accurately predict the electronic properties, the experimental phononic properties cannot be fully reproduced by a single unit-cell structure. 
In fact, the different orientations of the Si dimers contribute uniquely to the Raman spectrum of \SiTe{} \cite{chenAnisotropicOpticalProperties2020}, which we demonstrate by constructing two additional models with varying distributions of the vertical Si dumbbells. 
\begin{enumerate*}
    \item a layered phase with a vertical Si dumbbell in each layer, with \SI{50}{\percent} of all Si dumbbells oriented vertically;
    \item a phase where a vertical Si dumbbell in one layer is flipped to the horizontal orientation, referred to as the \SI{25}{\percent} phase (shown in Fig.~S5(a));
    \item a phase with no vertical Si dumbbells, such that all vertical Si dumbbells are flipped into the in-plane orientations, referred to as the 0\% vertical phase (shown in Fig.~S5(b)). 
\end{enumerate*}

The combined computed Raman spectra for these phases exhibit the three main experimental peaks (\cref{fig:layered-phase-properties}(a)) \cite{wangChemicallyTunableFull2018}.
Our results show that $\mathrm{A}_{\mathrm{1g}}^1$ mode is in the range from 124 to \SI{127}{cm^{-1}}, and $\mathrm{A}_{\mathrm{1g}}^2$ mode from 145 to \SI{148}{cm^{-1}}. In previous experiments, $\mathrm{A}_{\mathrm{1g}}^1$ was reported at \SI{127.0 \pm 0.4}{cm^{-1}}, while $\mathrm{A}_{\mathrm{1g}}^2$ was reported at \SI{143.9 \pm 0.2}{cm^{-1}} in \textcite{wangChemicallyTunableFull2018} and \SI{143.5 \pm 1.4}{cm^{-1}} in \textcite{johnson2019pressure}.
There is a 1--\SI{3}{\percent} error between theory and experiment that may be attributed to the approximated nature of the LDA functional.
From the resulting phonon normal modes, it is apparent that the $\mathrm{A}_{\mathrm{1g}}^1$ mode is sustained in layers where vertically oriented Si dimers are present 
(shown in Fig.~S6(a)).
Conversely, the $\mathrm{A}_{\mathrm{1g}}^2$ mode is sustained in layers without vertical dumbbells (shown in Fig.~S6(b)).
This indicates that the vibrational frequencies of the layers are affected by the fractional occupancy of vertical Si dumbbells.

Previous experiments were unsure whether the $\mathrm{E}_{\mathrm{g}}$ mode near \SI{140}{cm^{-1}} is inherent to \SiTe{} or arises from leakage of the $\mathrm{A}_{\mathrm{1g}}^2$ peak, since the sample may not sit entirely flush with the substrate \cite{wangChemicallyTunableFull2018}. 
Our theory predicts an $\mathrm{E}_{\mathrm{g}}$ mode with a frequency of \SI{138}{cm^{-1}} that corresponds to the in-plane vibration of the Te atoms combined with the stretching of the vertical Si dumbbells. Thus, this mode has an intrinsic physical origin but exists only when vertical dumbbells are included in the model. 
The $\mathrm{E}_{\mathrm{g}}$ Raman peak intensity calculated with 25\% vertical Si dumbbells is half that of the 50\% vertical structure, which also indicates that the $\mathrm{E}_{\mathrm{g}}$ mode is connected to the presence of vertical dimers.

Besides these three main peaks, we observed several groupings of peaks that share similar features.
At \qtyrange[range-phrase=--]{200}{206}{cm^{-1}}, all modes correspond to the translational motion of the horizontal Si dumbbells, where the frequencies are modulated by the dumbbell orientations. 
In \qtyrange[range-phrase=--]{305}{312}{cm^{-1}}, all modes correspond to the pure rotation of Si dimers where the rotation center is close to one of the Si atoms instead of the center of the dimers.
In \qtyrange[range-phrase=--]{386}{389}{cm^{-1}}, all modes correspond to the rotation of the Si dimers with an in-plane translational motion.
All modes in the range \qtyrange[range-phrase=--]{478}{492}{cm^{-1}} correspond to the stretching of Si dimers.
These modes are similar to those presented in a prior theoretical work of \textcite{bhattaraiAnisotropicOpticalProperties2020}.
We calculated very-low-intensity modes around \qtyrange[range-phrase=--]{95}{109}{cm^{-1}}, which is a similar frequency to that observed in previous Raman experiments, but the intensity is much less \cite{wangChemicallyTunableFull2018, zwick1976infrared}.
However, this frequency is near the band edge of the RazorEdge optical filter used in the Raman experiments and might suffer from instrumental effects in that region.

\subsubsection{Compression up to 7 GPa}

During compression from \SI{0}{GPa} to \SI{7}{GPa}, the lattice parameter is reduced in increments of \SI{0.16}{\angstrom}, scaling the cell volume from \SI{608.7}{\angstrom^3} to \SI{511.7}{\angstrom^3} and the interlayer distance $c$ from \SI{13.7}{\angstrom} to \SI{12.7}{\angstrom}.
The equation of state of the layered phase fitted with the Vinet equation of state is shown in \cref{fig:equation-of-states}(b).
The fitted values of the parameters in the Vinet equation of state are $K_o = \SI{20.65}{GPa}$, $K_o' = 7.73$, $V_o = \SI{609.65}{\angstrom^3}$, which agree well with those determined experimentally. 
 
Both theoretical and experimental Raman spectra \cite{johnson2019pressure} present a shift of the main peaks toward higher frequencies (\cref{fig:raman-compare-dos}(b)).
The experimental spectra also show suppression of the $\mathrm{A}_{\mathrm{1g}}^1$ peak as shown in (\cref{fig:raman-compare-dos}(a)).
Hence, the frequencies of the $\mathrm{A}_{\mathrm{1g}}$ modes are modulated by the thickness of the layers and interlayer separation.
In the theoretical Raman spectra, as pressure increases, the $\mathrm{A}_{\mathrm{1g}}^2$ peak is enhanced while the $\mathrm{A}_{\mathrm{1g}}^1$ peak is split into a doublet, which coincides with broadening the main peak and suppression of the left set shoulder in the experimental Raman spectra. 

The density of states in \cref{fig:raman-compare-dos}(c) shows that the indirect band gap drops from \SI{1.01}{eV} to \SI{0.23}{eV} when the pressure increases from \SI{0}{GPa} to \SI{7}{GPa}.
Meanwhile, the dielectric constant grows from 11.6 at \SI{0}{GPa} to 15.1 at \SI{7}{GPa}.
The trends of the density of states and dielectric constant imply that \SiTe{} becomes more metallic as the pressure increases, which is consistent with the suppressed intensity of Raman peaks at high pressure, as shown in \cref{fig:raman-compare-dos}(a).
This suggests that the material is metallic at pressures above \SI{7}{GPa}, as indicated by a previous experiment \cite{johnson2019pressure}. 

\subsubsection{Compression above 7 GPa and Pressure Release}
\label{sec:pressure_release}

\begin{figure}[t]
    \centering
    \includegraphics[width=\linewidth]{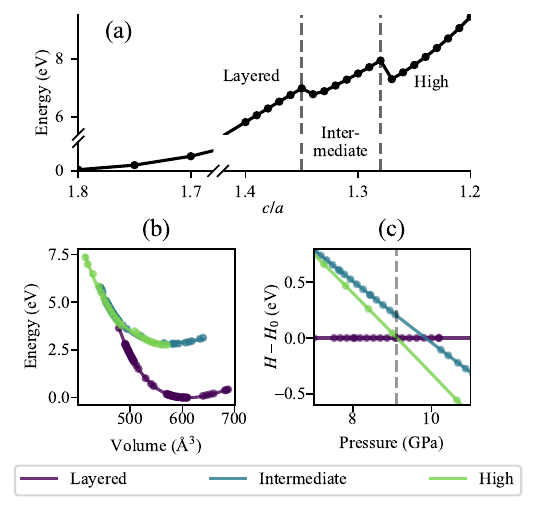}
    \caption{
    (a) Total energy of the various phases with respect to $c/a$, the ratio between the lattice parameters $c$ and $a$. Two dashed lines are the boundaries between different structural phases during the process of static compression. (b) Equation of state for the three \ce{Si2Te3} phases fitted with the Vinet equation of state. The zero point of the energy is set to the lowest energy of the layered phase. (c) The difference between the layered phase baseline enthalpy $H_0$ and the enthalpy of each phase $H$. Enthalpy of the high-pressure phase crosses that of the layered phase at \SI{9.1}{GPa}. 
    }
    \label{fig:equation-of-states}
\end{figure}


Before any structural phase transition occurs, \SiTe{} compresses more readily in the direction normal to the vdW layers (axis-$c$).
To promote a merger between layers, we simulate static compression by contracting $c/a$ (the length ratio between axis-$c$ and axis-$a$). This process induces the formation of bonds between layers and eventually a bulk structure.
The energy change in the process of the static compression is described in \cref{fig:equation-of-states}(a). 
A movie showing the structural change during the compression is provided in the SI. 
Atomic positions are fully relaxed at each compression step, and the total energy is carefully monitored to identify steps that undergo significant structural rearrangement.
When reducing $c/a$ from 1.85 to 1.2, \SiTe{} undergoes a two-step phase transition.
During the static compression, while $c/a$ remains larger than 1.35, the structure remains in the layered phase and the distance between layers is reduced. 
Reducing $c/a$ below 1.35 results in the formation of interlayer covalent bonds (shown in \cref{fig:equation-of-states}(a)), accompanied by a reorganization of the atomic geometry that reduces the pressure along axis $c$ from \SI{11.4}{GPa} to \SI{8.9}{GPa} and the total SCF energy by \SI{11}{meV/atom}.
This structure, which we call the intermediate phase, is maintained as $c/a$ and is further reduced to 1.28, achieving a pressure of \SI{10}{GPa}.
When $c/a$ reaches 1.27, corresponding to a pressure of \SI{10.8}{GPa}, the vdW stacks are entirely converted into a bulk phase, and the atomic geometry rearranges dramatically. 
The pairs of horizontal Si dumbbells rotate completely into the x-y plane, forming bonds horizontally with other Si dumbbells, and vertically aligned dumbbells connect between layers.
As shown in \cref{fig:equation-of-states}(a), this structural transition at $c/a=1.27$ reduces the total SCF energy by \SI{31}{meV/atom}.
We select the structure with $c/a=1.25$, which has a pressure of \SI{14}{GPa} but is not at hydrostatic conditions.
Cell parameters are optimized to a hydrostatic pressure of \SI{14}{GPa}.
We refer to the resulting bulk phase, shown in \cref{fig:phase-structure}(b), as the high-pressure phase.   

Decompression of the sample is simulated by incrementally scaling the lattice vectors by a constant factor.
The principle cell vector (axis-$a$) is enlarged from \SI{7.4}{\angstrom} to \SI{8.5}{\angstrom} in 40 steps, while axis-$b$ and axis-$c$ are scaled with the same proportions. 
Between each step atomic positions are relaxed to equilibrium, followed by the cell parameters under the pressure of the previous calculation.
The structure that achieves the lowest total energy is relaxed at \SI{0}{GPa} and hereafter referred to as the released phase.


\begin{figure*}
    \centering
    \includegraphics[width=\textwidth]{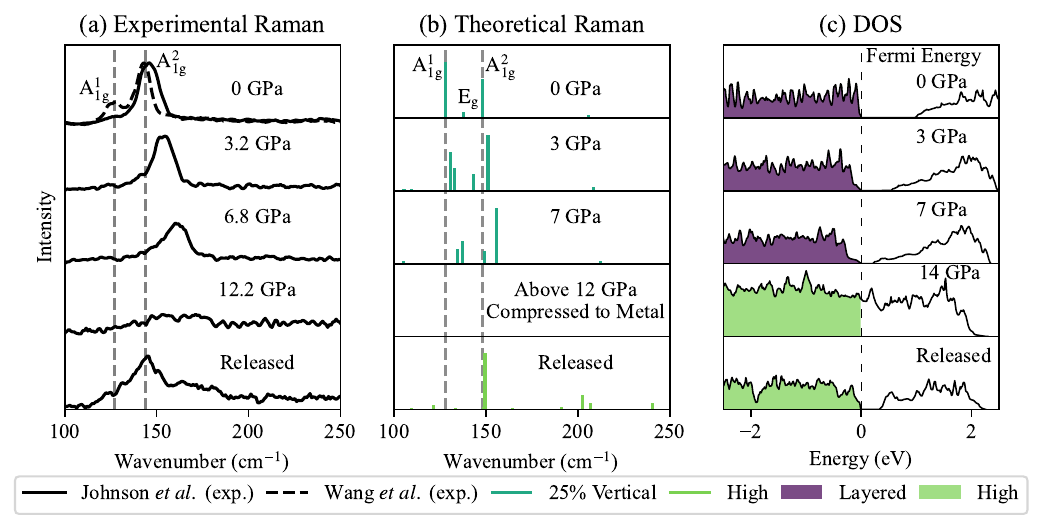}
    \caption{
    Experimental and theoretical Raman with calculated density of states under various pressures. (a) The experimental Raman spectra are reproduced from \textcite{wangChemicallyTunableFull2018} as dashed-line and \textcite{johnson2019pressure} as solid-line. (b) Raman spectra for several phases. The dashed lines are aligned to the Raman peaks of the 0 {GPa} phase. (c) Calculated density of states with the Fermi level aligned to 0 {eV}.
    }
    \label{fig:raman-compare-dos}
\end{figure*}

The band structure and density of states for the high-pressure phase are shown in Fig.~S4(b) and \cref{fig:raman-compare-dos}(c).
There is no band gap, indicating that the high-pressure phase is metallic.
This agrees with experiments, where no Raman response is detected (\cref{fig:raman-compare-dos}(a)), and the optical signature changes from transparent red to opaque grey/black under pressure.
The structure of the high-pressure phase is shown in \cref{fig:phase-structure}(b). 
There are 20 Si--Te bonds and 8 Si--Si bonds in a unit cell.
Compared with the layered phase, 4 Si--Te bonds break, and 4 Si--Si bonds form during the static compression. 
There are 12 Te atoms in the unit cell, 4 of which bond with 1 Si atom while the other 8 bond with 2 Si atoms. 
Si atoms always have between 3 and 6 neighbors.
There are eight Si atoms in the unit cell, two of which bond with one Si and two Te atoms, two bond with three Si and one Te atoms, two bond with two Si and 3 Te atoms, and two bond with 2 Si and 4 Te atoms. 
The average angles of Si--Si--Te and Si--Te--Si are \SI{89.9 \pm 11.9}{\degree} and \SI{92.5 \pm 10}{\degree}, which is well centered at \SI{90}{\degree}.
The structure is close to an octahedral geometry centered at Si atoms with many vacancies of the neighbors. 
This indicates that \SiTe{} converts from tetrahedral to octahedral coordination after compression. 

The same compression procedure and data analysis are repeated with the 25\% vertical phase and 0\% vertical phase, and the resulting structures are presented in Fig. S7. 
The local environment of Si atoms shows that the bond angles and bond lengths in both phases are similar to those found in the high-pressure phase.
The density of states indicates that also these high-pressure systems are metallic, even though the structures differ by local geometry.
We conclude that electronic properties and approximate structure can be captured by the high-pressure phase, but determining the underlying symmetry group is substantially complicated by disorder in the Si dimers. 

It is worth noting that the structure of the high-pressure phase shown in \cref{fig:phase-structure}(b) is strongly biased by the constraints of our first-principles modeling approach.
First, the initial orientation of the Si dimers will influence the direction in which silicon connects in the high-pressure phase.
Additionally, the small unit cell enforces a periodicity of the dimer orientation that the macroscopic sample does not possess.
Finally, our simulations occur at \SI{0}{K}, implying that the final or transition state configurations could be trapped by small metastable potential wells that would be overcome thermodynamically at ambient conditions.


The calculated equations of state for the high-pressure phase and intermediate phase are shown in \cref{fig:equation-of-states}(b) and (c). 
Fitted with the Vinet equation of state, the high-pressure phase has a bulk modulus of $K_o = \SI{12.5}{GPa}$ and a pressure derivative of the bulk modulus of $K_o' = 6.68$ and $V_o = \SI{585.7}{\angstrom^3}$, as shown in \cref{fig:equation-of-states}(b). 

In \cref{fig:equation-of-states}(c), the enthalpy of the high-pressure phase crosses the enthalpy of the layered phase at \SI{9.1}{GPa} which 
corresponds to the phase transition pressure measured in both past and current experiments \cite{johnson2019pressure}.
The enthalpy of the intermediate phase is never the lowest under any pressure indicating that it is not a thermodynamically stable phase.


The released phase structure is shown in \cref{fig:phase-structure}(c). 
Compared with the high-pressure phase, while most of the structural characteristics are maintained, there are two key changes.
First, 6 Si--Te bonds break, and 4 Si--Te bonds reform during the relaxation. 
This also impacts the bond coordination centered at Si atoms: 2 Si atoms bond with 1 Si and 3 Te atoms; 2 Si atoms bond with 3 Si and 2 Te atoms; 4 Si atoms bond with 2 Si and 2 Te atoms. 
Additionally, bond angles are slightly larger, which results in a more distorted octahedral coordination.
The average bond angles of Si--Si--Te and Te--Si--Te increase to \SI{97.55 \pm 9.0}{\degree} and \SI{104.3 \pm 9.6}{\degree}. 

The band structure and the density of states of the released phase are shown in Fig.~S4(c) and \cref{fig:raman-compare-dos}(c).
Noting that the released phase has a band gap, it indicates that \SiTe{} returns back to a semiconductor, which is consistent with the non-zero response of Raman spectrum from the experiment \cite{johnson2019pressure} as shown in \cref{fig:raman-compare-dos}(a). 
From the density of states, the indirect band gap of the released phase is \SI{0.36}{eV}, which is much smaller compared to the \SI{1.01}{eV} band gap of the layered phase. 
This means that the obtained released phase is more metallic than the initial \SiTe{} system at ambient conditions, which is consistent with the grey-red color in the optical images \cite{johnson2019pressure} of \SiTe{} after the pressure release. 

The theoretically predicted Raman spectrum of the released phase is shown in \cref{fig:raman-compare-dos}(b) which has an excellent agreement with the experimental Raman spectrum of \textcite{johnson2019pressure} as shown in \cref{fig:raman-compare-dos}(a). 
Compared with the theoretical Raman spectrum of \SiTe{} at \SI{0}{GPa}, the initial main peak at \SI{148}{cm^{-1}} in the Raman spectrum returns back to \SI{149}{cm^{-1}} after the pressure is released, which has the same trend as the experiment found initially at \SI{143.9}{cm^{-1}} and returning back to \SI{143.4}{cm^{-1}}. 

\subsubsection{Model Validity and the Influence of Occupational Disorder}

Although the theoretical treatment of only three phases cannot capture the properties of occupational disorder, the electronic and phononic properties remain similar to experimental values, both qualitatively and quantitatively.
Numerous theoretical treatments of the \SiTe{} system in the past decade \cite{shenVariabilityStructuralElectronic2016,chenAnisotropicOpticalProperties2020,bhattaraiCOMPUTATIONALSTUDYOPTICAL2021,steinberg2016search,RevisitingDronskowski2017,uzhhorod,juneja2017high,kimFirstPrinciplesInvestigationsSemiconductortoMetal2023} present different structures, but still agree on several calculated properties.
In our study, the density of states for the three arrangements of Si dumbbells at ambient conditions (\cref{fig:layered-phase-properties}(b)) do not have significant qualitative differences.
The calculated indirect band gaps for the layered phase, 25\% vertical phase and 0\% vertical phase are \SI{1.01}{eV}, \SI{0.88}{eV}, and \SI{1.01}{eV}, respectively.
All are within the uncertainty of past UV--Vis absorption experiments that report an indirect band gap of about \SI{1}{eV} \cite{wangChemicallyTunableFull2018}.
A detailed discussion of the experimental analysis is presented in the supporting information. 

Two mysteries remain in the observed properties of \SiTe{}.
The first is the mechanism of the defect trap states that can be populated by illuminating the sample with band gap energy photons \cite{ziegler1977photoelectric,wangChemicallyTunableFull2018,chenAnisotropicOpticalProperties2020}, which we attribute to a spectrum of closely spaced momentum-indirect exciton states originating from dimer disorder and exhibiting phonon-assisted photoluminescence \cite{brem2020phonon_assisted}.
Second, the energetic ground state configuration of the layered phase requires \SI{50}{\percent} of the Si dimers to orient vertically, while XRD measurements indicate that only \SI{23.5}{\percent} Si dimers are vertically aligned \cite{ploog1976crystal,keuleyan2015silicon}.
VLS synthesis techniques cause precipitation of nanoplates or nanocrystals from a droplet of liquid Te and Si \cite{keuleyan2015silicon}. 
Si dimer orientations are governed by kinetics at the liquid-solid interface during synthesis and become thermodynamically locked as the layers grow.
In our model, the energy difference between the 50\% vertical and 0\% vertical configurations is \SI{12.5}{meV/atom}, indicating that the energy difference between the two orientations is less than the equipartition kinetic energy at ambient conditions.
Transitioning a Si dimer from one orientation to another after synthesis, however, would require the redistribution of covalent electrons and a relatively high energy barrier.
\textcite{shenVariabilityStructuralElectronic2016} find from \textit{ab initio} nudged elastic band (NEB) calculations that dimer orientations are separated by an energy barrier of approximately \SI{1}{eV}, preventing dimers from changing configuration at the ambient temperature.

\section{Conclusion}

This work elucidates the behavior of \SiTe{} under high pressure through first-principles calculations and experiments.
A low-pressure phase transition ($<\SI{1}{GPa}$) is found coincident with a trigonal to hexagonal change in the crystal structure.
The hexagonal structure is maintained up to \SI{8}{GPa}.
No amorphization is seen up to \SI{11.5}{GPa} in these experiments, though new peaks are found in the XRD above \SI{8.5}{GPa}.
First-principles calculations capture the metallization of silicon telluride at high pressure (above \SI{9.1}{GPa}) as seen in previous Raman experiments of \textcite{johnson2019pressure}, and characteristic Raman signatures are attributed to the orientation of Si dimers using DFPT calculations.
Because Raman modes are sensitive to the Si dimer orientation, we construct a three-phase model to explain the experimental Raman spectra at lower pressure.
At high pressure, the Raman spectrum agrees with our experimental results due to metallization and is not dramatically influenced by the initial dimer orientation.
The most complicated, and unresolved, task is to explain the structural properties of the high-pressure phase in terms of the experimental XRD since partial occupancy and finite size effects strongly influence the resulting geometry.
There is no evident change in \ce{Si2Te3} stoichiometry with pressure, contrary to previous theoretical studies \cite{grzechnik2022chemical, steinberg2016search}.
Our investigation reveals interesting trends in both electronic and vibrational properties of \SiTe{} under hydrostatic compression and includes an accurate theoretical description of the structural phase transition observed experimentally at \SI{9.1}{GPa}.

\begin{acknowledgments}
The authors acknowledge funding from the National Science Foundation NSF-DMR-2202472. The theoretical work was supported by the U.S. Department of Energy, Office of Basic Energy Sciences, Division of Materials Science and Engineering (Grant No. DE-SC0022288).
This research used resources from the Advanced Light Source, which is a U.S. Department of Energy Office of Science User Facility under contract no. DE-AC02-05CH11231.
\end{acknowledgments}

\bibliography{References}

\end{document}